\documentclass[letterpaper, 10pt, conference]{ieeeconf}      

\IEEEoverridecommandlockouts                              
\usepackage{amsmath} 
\usepackage{amssymb}  

\usepackage{cite}
\usepackage{amsfonts}
\usepackage{graphicx}
\usepackage{textcomp}
\usepackage{xcolor}
\usepackage{lipsum}
\usepackage{algorithm}
\usepackage{algpseudocode}

\newtheorem{assumption}{Assumption}
\newtheorem{definition}{Definition}

\title{\LARGE \bf
A Dynamic Generalized Kalman Consensus Filter for Switching\\Sensor Networks*
}

\author{Tirthankar Chakraborty and Arnab Maity
\thanks{*Published in the Proceedings of the 2026 15th Asian Control Conference (ASCC). DOI: 10.1109/ASCC70886.2026.11702283.}
\thanks{© 2026 IEEE. Personal use of this material is permitted. Permission from IEEE must be obtained for all other uses, in any current or future media, including reprinting/republishing this material for advertising or promotional purposes, creating new collective works, for resale or redistribution to servers or lists, or reuse of any copyrighted component of this work in other works.}
\thanks{The authors are affiliated with the Department of Aerospace Engineering at the Indian Institute of Technology Bombay, Mumbai 400076, India. Emails: Tirthankar Chakraborty (tirthankar@aero.iitb.ac.in), Arnab Maity (arnab@aero.iitb.ac.in).}%
}

\begin{document}

\maketitle
\thispagestyle{empty}
\pagestyle{empty}

\begin{abstract}

Distributed state estimation is critical for applications such as surveillance, autonomous navigation, and wide-area monitoring, where sensor agents must cooperatively track targets using only local measurements and neighbor-to-neighbor communication. Existing distributed filters have been shown to achieve accurate estimation even under sparse inter-agent communication and limited sensing ranges. However, many of these methods rely on consensus parameters that depend on global properties of the communication graph, such as the maximum degree of the graph, and are therefore sensitive to changes in network topology. This limitation is particularly significant in sensor networks with mobile agents, where communication links change over time. This paper presents a Dynamic Generalized Kalman Consensus Filter for target tracking in sensor networks with switching communication topologies. The proposed algorithm computes information-based consensus weights using only locally available quantities, eliminating the need for global network parameters. Numerical simulations demonstrate that the proposed algorithm maintains estimation accuracy under switching network topologies and outperforms existing distributed filters in the given tracking problem.

\end{abstract}


\section{INTRODUCTION}

Distributed sensor networks, composed of spatially dispersed but interconnected sensing agents, have become a critical tool for monitoring complex, dynamic environments. Through cooperative behavior among individual agents, such networks can collectively track dynamic processes with improved robustness and scalability. Applications span a wide spectrum, from cooperative reconnaissance and infrastructure monitoring to autonomous navigation, surveillance, and environmental monitoring, highlighting their growing importance across diverse domains \cite{he2020distributed, modalavalasa2021review}.

Estimating the state of dynamic processes or targets within a noisy sensing environment is a fundamental challenge in sensor networks \cite{anderson2005optimal}. For over six decades, the Kalman filter \cite{kalman1960new} has served as a powerful and widely used algorithm for addressing this problem. A \textit{centralized} approach requires all sensor agents to transmit their observations to a fusion center for processing. However, limited communication bandwidth and processing constraints render this approach impractical as the network size grows. Furthermore, routing all data through a single node introduces a critical vulnerability to failure.

A decentralized Kalman filter requiring all-to-all connectivity was presented in \cite{rao1993fully}, which improved robustness but did not address scalability. The first scalable distributed Kalman filter was introduced in \cite{olfati2005distributed}, utilizing embedded \textit{average consensus} filters, though its performance was found to be inadequate in subsequent studies \cite{olfati2007distributed}. The Kalman Consensus Filter (KCF) was later introduced in \cite{olfati2009kalman} as another scalable distributed filtering algorithm widely used in target tracking. However, in scenarios where some agents in the network are \textit{oblivious} (i.e., lack measurement data), as is common in large-area networks with agents of limited sensing range, the performance of KCF was found to be inadequate.

This deficiency in KCF originates from its reliance on \textit{unweighted} average consensus, in which every agent assigns equal weight to information received from neighboring agents, including oblivious ones. It has been demonstrated that this can lead to divergence \cite{khan2019optimal}. Several algorithms were subsequently proposed that employ \textit{weighted} average consensus to address this deficiency, including the Generalized Kalman Consensus Filter (GKCF) \cite{kamal2011generalized} and the Information-weighted Consensus Filter (ICF) \cite{kamal2013information}, among others \cite{hu2011diffusion,battistelli2014consensus}.

A dynamic active average consensus algorithm was proposed in \cite{chung2020dynamic,peterson2019active} for distributed control applications. There, the agents with access to measurements are designated \textit{active} agents, and the objective is for all agents to asymptotically track the mean of the active agents' measurements. To address the suboptimality induced by oblivious agents, the Approximate Distributed Kalman Filter (ADKF) was proposed in \cite{battilotti2020asymptotically} for continuous-time systems, with its estimation error shown to converge asymptotically to that of the centralized Kalman filter (CKF). Furthermore, \cite{battilotti2021stability} established that for discrete-time systems, achieving near-optimal performance demands a sufficient number of consensus iterations.

However, a key limitation of most existing distributed filtering methods employing consensus protocols is the assumption of a fixed communication topology. This assumption is often violated in practice because mobile agents may move in and out of communication range, leading to a switching communication topology. In such cases, the inability of these filters to adapt to a dynamic network can degrade performance, particularly in the presence of oblivious agents or intermittent observations. This creates a requirement for a consensus-based distributed filtering method capable of maintaining estimation accuracy under simultaneously switching communication topologies and heterogeneous agent observability.

The primary contribution of this paper is the development of the Dynamic Generalized Kalman Consensus Filter (DGKCF) for distributed target tracking in networks with switching communication topologies. This algorithm employs an embedded weighted average consensus protocol and uses locally computed information-based consensus weights, without the knowledge of any global network parameter. Numerical simulations demonstrate that the proposed algorithm achieves lower root mean square error (RMSE), lower mean absolute error (MAE), and faster convergence than the other distributed filters under consideration, particularly in scenarios involving oblivious agents and switching communication topologies. Additionally, the proposed algorithm has a computational cost comparable to that of other distributed algorithms that use weighted consensus protocols.

The rest of the paper is structured as follows. Section II outlines the problem formulation. Section III develops the proposed algorithm. Section IV presents the simulation results for a target-tracking scenario, and Section V provides the conclusion of the paper.

In this paper, scalars are represented by lowercase letters, vectors by bold lowercase letters, and matrices by bold uppercase letters. We denote the transpose by $(\cdot)^\top$ and the stochastic expectation by $\mathbb{E}[\cdot]$. $\mathbb{Z}_{\ge0}$ denotes the set of positive integers.

\section{PROBLEM FORMULATION}

The state evolution of the system is modeled by a discrete-time, linear time-invariant (LTI) dynamical system. A network of $N$ sensor agents is used to monitor the system. The system dynamics are described as follows:
\begin{align}\label{sys1}
    \mathbf{x}(k+1) &= \mathbf{A}\mathbf{x}(k) + \mathbf{B}\mathbf{w}(k),
\end{align}
where $\mathbf{x}(k) \in \mathbb{R}^{n}$ denotes the system state, $\mathbf{A} \in \mathbb{R}^{n \times n}$ denotes the state propagation matrix, $\mathbf{w}(k) \in \mathbb{R}^{n}$ represents the process noise, and $\mathbf{B} \in \mathbb{R}^{n \times n}$ is the process noise input matrix. The measurement model for agent $i$ is given by
\begin{align}\label{meas1}
    \mathbf{z}_i(k) &= \mathbf{H}_i\mathbf{x}(k) + \mathbf{v}_i(k),\ i \in \{ 1, 2, \ldots, N\},
\end{align}
where $\mathbf{z}_i(k) \in \mathbb{R}^{m}$ is the measurement vector, $\mathbf{H}_i \in \mathbb{R}^{m\times n}$ is the measurement matrix, and $\mathbf{v}_i(k) \in \mathbb{R}^{m}$ is the measurement noise of agent $i$. The vectors $\mathbf{w}(k)$ and $\mathbf{v}_i(k)$ are modeled as zero-mean white Gaussian processes with $\mathbf{Q}$ and $\mathbf{R}_i$ being their respective covariance matrices.

\begin{assumption}
The initial state, measurement noise, and process noise are mutually independent.
\end{assumption}

The overall system state is assumed to be observable when considering all agents in the network collectively, even though individual agents are not required to observe the entire state on their own.

The sensor network is represented by an undirected graph $G_{p(k)} = (\mathcal{V}, \mathcal{E}_{p(k)})$, where elements of the vertex set $\mathcal{V} = \{1, 2, \ldots, N\}$ denote the agents, and elements of the edge set $\mathcal{E}_{p(k)} \subseteq \mathcal{V} \times \mathcal{V}$ denote the communication links. The set of admissible communication topologies is defined as $\mathcal{G} \triangleq \{G_1,\ldots, G_h\}$. A piecewise constant switching signal $p(k):\mathbb{Z}_{\ge0}\to\{1,\ldots,h\}$ selects the active topology $G_{p(k)}\in\mathcal{G}$. 

\begin{assumption}
Every graph $G_l \in \mathcal{G}$ is connected. That is, every admissible graph contains an undirected path between any pair of vertices.
\end{assumption}

Let $\mathcal{Z}_i(k) = \{\mathbf{z}_i(0),\mathbf{z}_i(1),\ldots,\mathbf{z}_i(k)\}$ denote the collection of measurements available up to time $k$. The state estimates at agent $i$ are then given by
\begin{align}
    \bar{\mathbf{x}}_i(k) &= \mathbb{E}[\mathbf{x}(k) | \mathcal{Z}_i(k-1)],\\
    \hat{\mathbf{x}}_i(k) &= \mathbb{E}[\mathbf{x}(k) | \mathcal{Z}_i(k)],
\end{align}
where $\bar{\mathbf{x}}_i(k)$ denotes the \textit{prior} (predicted) estimate and $\hat{\mathbf{x}}_i(k)$ denotes the \textit{posterior} (updated) estimate of the state vector $\mathbf{x}(k)$ at agent $i$, respectively. The prior estimation error is defined as $\bar{\boldsymbol{\eta}}_i(k) = \bar{\mathbf{x}}_i(k) -\mathbf{x}(k)$ and the posterior estimation error is defined as $\hat{\boldsymbol{\eta}}_i(k) = \hat{\mathbf{x}}_i(k) -\mathbf{x}(k)$. The prior and posterior error covariance matrices for these estimates are as follows:
\begin{align}
    \mathbf{P}_i(k) &= \mathbb{E}\left[ \bar{\boldsymbol{\eta}}_i(k) \bar{\boldsymbol{\eta}}_i^\top(k) \right],\\
    \mathbf{M}_i(k) &= \mathbb{E}\left[ \hat{\boldsymbol{\eta}}_i(k) \hat{\boldsymbol{\eta}}_i^\top(k) \right].
\end{align}
The primary goal is to develop a local estimator at each agent that produces an accurate estimate $\hat{\mathbf{x}}_i(k)$ of the system state $\mathbf{x}(k)$. The Kalman filter \cite{kalman1960new} is a linear estimator whose posterior state update equation at agent $i$ is given by
\begin{align}
    \hat{\mathbf{x}}_i(k) = \bar{\mathbf{x}}_i(k) + \mathbf{K}_i(k)\left[ \mathbf{z}_i(k) - \mathbf{H}_i\bar{\mathbf{x}}_i(k) \right],
\end{align}
where $\mathbf{K}_i(k) \in \mathbb{R}^{n \times m}$ is the Kalman gain. To highlight the significance of agent indices, we use a simplified notation throughout this paper that excludes the time index $k$ from all estimator expressions. Using this approach, the estimator above can be expressed as follows:
\begin{align}
    \hat{\mathbf{x}}_i = \bar{\mathbf{x}}_i + \mathbf{K}_i\left( \mathbf{z}_i - \mathbf{H}_i\bar{\mathbf{x}}_i \right).
\end{align}
Furthermore, the notation $\mathbf{x}^+$ denotes the value of a variable $\mathbf{x}$ at the next time step. For example, $\mathbf{x}^+ = f(\mathbf{x}, \mathbf{w})$ is the compact form of the expression $\mathbf{x}(k+1) = f\left( \mathbf{x}(k), \mathbf{w}(k) \right)$.


\section{ALGORITHM DEVELOPMENT}

The Kalman filter computes the posterior state estimate at each time step by fusing the prior estimate with the latest measurement data. The Kalman Consensus Filter (KCF) \cite{olfati2009kalman} extends this framework by incorporating average consensus, enabling distributed state estimation across a network of sensor agents. The posterior state update equation of KCF at agent $i$ is given by
\begin{align}
    \hat{\mathbf{x}}_i = \bar{\mathbf{x}}_i + \mathbf{K}_i\left( \mathbf{z}_i - \mathbf{H}_i\bar{\mathbf{x}}_i \right) + \mathbf{C}_i \sum_{j \in \mathcal{N}_i} (\bar{\mathbf{x}}_j - \bar{\mathbf{x}}_i),
\end{align}
where $\mathbf{C}_i \in \mathbb{R}^{n \times n}$ denotes the consensus gain and the set of neighboring agents of agent $i$ is denoted by $\mathcal{N}_i$. This filter can be expressed in the \textit{information form} \cite{olfati2009kalman} using the fused measurement information vector $\mathbf{y}_i$ and the fused measurement information matrix $\mathbf{S}_i$ as
\begin{align}
    \hat{\mathbf{x}}_i = \bar{\mathbf{x}}_i + \mathbf{M}_i\left( \mathbf{y}_i - \mathbf{S}_i\bar{\mathbf{x}}_i \right) + \mathbf{C}_i \sum_{j \in \mathcal{N}_i} (\bar{\mathbf{x}}_j - \bar{\mathbf{x}}_i),
\end{align}
where $\mathbf{y}_i$ and $\mathbf{S}_i$ are defined as follows:
\begin{align}
    \mathbf{y}_i &= \sum_{j \in \mathcal{J}_i} \mathbf{H}_j^\top \mathbf{R}_j^{-1} \mathbf{z}_j, \label{def_yi} \\
    \mathbf{S}_i &= \sum_{j \in \mathcal{J}_i} \mathbf{H}_j^\top \mathbf{R}_j^{-1} \mathbf{H}_j. \label{def_Si}
\end{align}
Here, $\mathcal{J}_i$ denotes the \textit{inclusive neighborhood} of agent $i$, i.e., the union of agent $i$ and the set of neighbors of agent $i$.

\begin{figure}[ht!]
    \centering
    \includegraphics[width=0.45\textwidth]{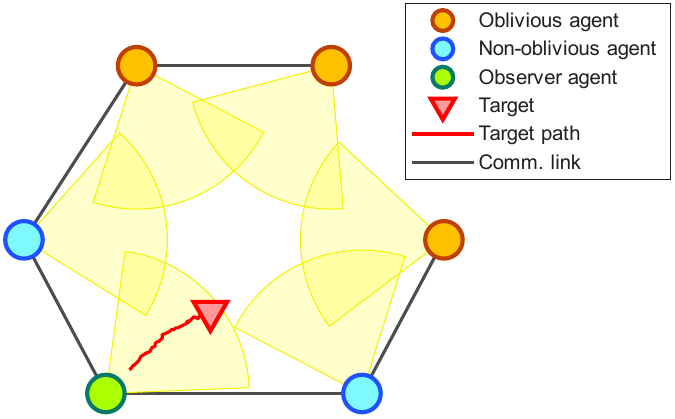}
    \caption{Illustration of oblivious agents in a sensor network.}
    \label{oblivious_agents}
\end{figure}

In wide-area distributed sensor networks with agents that have limited sensing range and sparse communication topologies, as illustrated in Fig.~\ref{oblivious_agents}, some agents may have no access to current measurement information.

\begin{definition}[Oblivious Agent]
    Let $\mathcal{D} \subseteq \mathcal{V}$ denote the set of agents directly observing the target. An agent $i$ is \emph{oblivious} if it belongs to the set
    \begin{align*}
        \mathcal{O} = \{i \in \mathcal{V} \mid i \notin \mathcal{D},\ \mathcal{N}_i \cap \mathcal{D} = \varnothing  \}.
    \end{align*}
\end{definition}

In other words, an oblivious agent neither observes the target directly nor has any neighboring agents that do. In sparse networks, oblivious agents can experience considerable delays in updating their estimates. Such delays arise because these agents lack direct access to current target observations and must instead rely on the state estimates of neighboring agents. Since these neighbors last received updates in prior iterations, the information propagated to an oblivious agent may already be outdated. As a result, an oblivious agent may lag several iterations behind in acquiring current information about the target.

Since the KCF employs average consensus, the estimates from oblivious agents, which are often highly inaccurate, are assigned the same weight as those from observing agents. Such equal weighting degrades the estimation performance of the filter. The complete KCF operation is summarized in Algorithm \ref{alg:kcf}.


\begin{algorithm}
\caption{Kalman Consensus Filter}\label{alg:kcf}
\begin{algorithmic}

\State \textbf{Given:} $\mathbf{P}_i(0) = \mathbf{P}_0$, $\bar{\mathbf{x}}_i(0) = \mathbf{x}_0$, $\mathcal{J}_i = \mathcal{N}_i \cup \{i\}$.\\
At time step $k$, sensor agent $i$,

\State 1: Collect observation $\mathbf{z}_i$.

\State 2: Calculate local information matrix and vector:
\begin{align*}
    \quad \mathbf{U}_i &= \mathbf{H}_i^\top \mathbf{R}_i^{-1} \mathbf{H}_i,\\
    \mathbf{u}_i &= \mathbf{H}_i^\top \mathbf{R}_i^{-1} \mathbf{z}_i.
\end{align*}

\State 3: Transmit $\{\mathbf{U}_i, \mathbf{u}_i, \bar{\mathbf{x}}_i \}$ to all neighboring agents.

\State 4: Obtain $\{\mathbf{U}_j, \mathbf{u}_j, \bar{\mathbf{x}}_j \}$ from neighboring agents.

\State 5: Fuse information matrices and vectors:
\begin{align*}
    \quad \mathbf{S}_i &= \sum_{j \in \mathcal{J}_i} \mathbf{U}_j,\\
    \mathbf{y}_i &= \sum_{j \in \mathcal{J}_i} \mathbf{u}_j.
\end{align*}

\State 6: Calculate the updated estimate:
\begin{align*}
    \gamma &= \dfrac{\epsilon}{\| \mathbf{P}_i \|_F + 1}, \quad \| \mathbf{P}_i \|_F = \sqrt{\operatorname{tr}(\mathbf{P}_i^\top \mathbf{P}_i)}, \\[1ex]
    \mathbf{M}_i &= \left( \mathbf{S}_i + \mathbf{P}^{-1}_i \right)^{-1}, \\[1ex]
    \hat{\mathbf{x}}_i &= \bar{\mathbf{x}}_i + \mathbf{M}_i \left( \mathbf{y}_i - \mathbf{S}_i \bar{\mathbf{x}}_i \right) + \gamma \mathbf{P}_i \sum_{j \in \mathcal{N}_i} (\bar{\mathbf{x}}_j - \bar{\mathbf{x}}_i).
\end{align*}

\State 7: Propagate the local state estimate in time:
\begin{align*}
    \mathbf{P}_i^+ &\leftarrow \mathbf{A} \mathbf{M}_i \mathbf{A}^\top + \mathbf{B} \mathbf{Q} \mathbf{B}^\top,\\
    \quad \bar{\mathbf{x}}_i^+ &\leftarrow \mathbf{A} \hat{\mathbf{x}}_i.
\end{align*}
\end{algorithmic}
\end{algorithm}


To address the issues encountered by the KCF in the presence of oblivious agents, several methods, including the Generalized Kalman Consensus Filter (GKCF) \cite{kamal2011generalized}, that incorporate a weighted average consensus protocol were proposed. The term \textit{weight} here refers to the prior information matrix, i.e., the inverse of the prior covariance matrix. Since the prior covariance matrix ($\mathbf{P}_i$) is a measure of uncertainty in the prior estimate ($\bar{\mathbf{x}}_i$), the inverse of this matrix ($\mathbf{P}_i^{-1}$) is a measure of the confidence in the prior estimate. The weighted average consensus update is then given by
\begin{align}
    \widetilde{\mathbf{x}}_i &= \widetilde{\mathbf{P}}_i  \left(\mathbf{P}_i^{-1}\bar{\mathbf{x}}_i + \epsilon \sum_{j \in \mathcal{N}_i} \left( \mathbf{P}_j^{-1}\bar{\mathbf{x}}_j - \mathbf{P}_i^{-1}\bar{\mathbf{x}}_i \right) \right),
\end{align}
where $\widetilde{\mathbf{x}}_i$ is the weighted average consensus estimate and $\epsilon$ is the consensus step size, which is a parameter satisfying $0 < \epsilon < 1/\Delta$, with $\Delta$ being the maximum degree of the communication network graph. $\widetilde{\mathbf{P}}_i$ is the result of average consensus on the weights and is defined as
\begin{align} \label{tildeP}
    \widetilde{\mathbf{P}}_i^{-1} &= \mathbf{P}_i^{-1} + \epsilon \sum_{j \in \mathcal{N}_i} \left( \mathbf{P}_j^{-1} - \mathbf{P}_i^{-1} \right).
\end{align}

As the state estimates converge toward the weighted average, the associated weights simultaneously approach their average. This step ensures that the weights are properly updated when an oblivious agent receives more informative state estimates.

The consensus step size is a function of the maximum degree of the communication graph ($\Delta$). For faster convergence, this parameter should be chosen close to $1/\Delta$, without exceeding it. Exceeding this limit would result in the divergence of estimates. In most existing filters, this parameter is fixed before the filtering process, and there is no mechanism for its online adaptation. This introduces a key limitation: as the communication graph and its maximum degree change, the fixed $\epsilon$ may become suboptimal, resulting in either slow convergence or divergence of estimates.


To address this limitation, this paper proposes the Dynamic Generalized Kalman Consensus Filter (DGKCF), which replaces the fixed consensus step size with an information-based scalar weight. Here, the weighted average consensus protocol is reformulated to reflect the relative quality of prior estimates across agents. 

Let the information value related to agent $i$ and agent $j$ be defined as
\begin{align}
\phi_{ij} = \dfrac{1}{\operatorname{tr}(\mathbf{P}_i)} + \dfrac{1}{\operatorname{tr}(\mathbf{P}_j)},
\end{align}
where $\operatorname{tr}(\mathbf{P})$ denotes the trace of $\mathbf{P}$. The term $\operatorname{tr}(\mathbf{P}_i)$ is the sum of the variances of the state components and provides a scalar measure of the total estimation uncertainty at agent $i$. Its inverse, $1/\operatorname{tr}(\mathbf{P}_i)$, converts this uncertainty measure into a scalar indicator of information content, assigning larger values to more reliable estimates. By symmetrically combining the information measures of both agents $i$ and $j$, $\phi_{ij}$ describes the mutual reliability associated with the communication link between them.

The trace is preferred over alternative scalar measures, such as the matrix determinant or spectral norm, as it is computationally efficient (requiring only a diagonal sum) and numerically well-conditioned.

Then, for every agent $j \in \mathcal{J}_i$, the normalized consensus weight is given by
\begin{align}\label{weights}
    w_{ij} = \dfrac{\phi_{ij}}{\sum_{l \in \mathcal{J}_i} \phi_{il}} \ .
\end{align}
Consequently, neighboring agents with lower uncertainty receive larger normalized consensus weights and exert greater influence on the fused estimate. In the limiting case where all agents share equal covariances, the weights reduce to $w_{ij} = 1/|\mathcal{J}_i|$. By construction, the weights satisfy the following properties:
\begin{align*}
    w_{ij} \ge 0,
    \qquad
    \sum_{j \in \mathcal{J}_i} w_{ij} = 1.
\end{align*}

The weighted average consensus estimate for agent $i$ is then computed as
\begin{align}
    \widetilde{\mathbf{P}}^{-1}_i &= \mathbf{P}_i^{-1} + \sum_{j \in \mathcal{N}_i} w_{ij} \left( \mathbf{P}_j^{-1} - \mathbf{P}_i^{-1} \right), \label{tildePwij}\\
    \widetilde{\mathbf{P}}_i^{-1} \widetilde{\mathbf{x}}_i &= \mathbf{P}_i^{-1}\bar{\mathbf{x}}_i + \sum_{j \in \mathcal{N}_i} w_{ij} \left( \mathbf{P}_j^{-1}\bar{\mathbf{x}}_j - \mathbf{P}_i^{-1}\bar{\mathbf{x}}_i \right).
\end{align}

Since the weights are nonnegative and sum to one over the inclusive neighborhood, the consensus update preserves the convexity properties required for stable information fusion. When additional communication is allowed, multiple weighted consensus iterations may be carried out between measurement updates so that local estimates converge further toward the global information-weighted average.

The consensus estimate $\widetilde{\mathbf{x}}_i$ is then refined using locally available measurement information as
\begin{align}
    \hat{\mathbf{x}}_i &= \widetilde{\mathbf{x}}_i + \mathbf{M}_i \left( \mathbf{y}_i - \mathbf{S}_i \widetilde{\mathbf{x}}_i \right),\\
    \mathbf{M}_i &= \left( \mathbf{S}_i + \widetilde{\mathbf{P}}^{-1}_i \right)^{-1},
\end{align}
where the fused measurement information vector $\mathbf{y}_i$ and matrix $\mathbf{S}_i$ are as defined in \eqref{def_yi} and \eqref{def_Si}, respectively. The complete DGKCF procedure is summarized in Algorithm \ref{alg:dgkcf}.

The proposed weighting scheme possesses three important properties. First, it can handle the presence of oblivious agents in the network by assigning larger consensus weights to more reliable (lower uncertainty) estimates and smaller weights to less reliable ones. Second, it is fully distributed, as all weights are computed using only locally available covariance data and information received from neighboring agents, without requiring any global knowledge of the network. Third, the weights are recomputed at every time step, enabling the consensus process to adapt automatically to changes in both network connectivity and the relative quality of the local estimates.

Several existing approaches also incorporate covariance information into the consensus process. In contrast, the proposed algorithm derives normalized weights directly from a local information metric and does not require a network-dependent convergence parameter. This eliminates the need for selecting a step size based on the maximum graph degree and enables the filter to adapt to switching network topologies.


\begin{algorithm}
\caption{Dynamic Generalized Kalman Consensus Filter}\label{alg:dgkcf}
\begin{algorithmic}

\State \textbf{Given:} $\mathbf{P}_i(0) = \mathbf{P}_0$, $\bar{\mathbf{x}}_i(0) = \mathbf{x}_0$, $\mathcal{J}_i = \mathcal{N}_i \cup \{i\}$.\\
At time step $k$, sensor agent $i$,

\State 1: Collect observation $\mathbf{z}_i$.

\State 2: Calculate local information matrix and vector:
\begin{align*}
    \quad \mathbf{U}_i &= \mathbf{H}_i^\top \mathbf{R}_i^{-1} \mathbf{H}_i,\\
    \mathbf{u}_i &= \mathbf{H}_i^\top \mathbf{R}_i^{-1} \mathbf{z}_i.
\end{align*}

\State 3: Transmit $\{\mathbf{U}_i, \mathbf{u}_i, \bar{\mathbf{x}}_i, \mathbf{P}_i \}$ to all neighboring agents.

\State 4: Obtain $\{\mathbf{U}_j, \mathbf{u}_j, \bar{\mathbf{x}}_j, \mathbf{P}_j \}$ from neighboring agents.

\State 5: Fuse information matrices and vectors:
\begin{align*}
    \quad \mathbf{S}_i &= \sum_{j \in \mathcal{J}_i} \mathbf{U}_j,\\
    \mathbf{y}_i &= \sum_{j \in \mathcal{J}_i} \mathbf{u}_j.
\end{align*}

\State 6: Calculate weighted average consensus estimate:
\begin{align*}
    \phi_{ij} &= \dfrac{1}{\operatorname{tr}(\mathbf{P}_i)} + \dfrac{1}{\operatorname{tr}(\mathbf{P}_j)},\\
    w_{ij} &= \dfrac{\phi_{ij}}{\sum_{l \in \mathcal{J}_i} \phi_{il}}\ ,\ \forall j \in \mathcal{J}_i,\\[1ex]
    \widetilde{\mathbf{P}}^{-1}_i &= \mathbf{P}_i^{-1} + \sum_{j \in \mathcal{N}_i} w_{ij} \left( \mathbf{P}_j^{-1} - \mathbf{P}_i^{-1} \right),\\
    \widetilde{\mathbf{P}}_i^{-1} \widetilde{\mathbf{x}}_i &= \mathbf{P}_i^{-1}\bar{\mathbf{x}}_i + \sum_{j \in \mathcal{N}_i} w_{ij} \left( \mathbf{P}_j^{-1}\bar{\mathbf{x}}_j - \mathbf{P}_i^{-1}\bar{\mathbf{x}}_i \right).
\end{align*}

\State 7: Calculate the updated estimate:
\begin{align*}
    \mathbf{M}_i &= \left( \mathbf{S}_i + \widetilde{\mathbf{P}}^{-1}_i \right)^{-1}, \\
    \hat{\mathbf{x}}_i &= \widetilde{\mathbf{x}}_i + \mathbf{M}_i \left( \mathbf{y}_i - \mathbf{S}_i \widetilde{\mathbf{x}}_i \right).
\end{align*}

\State 8: Propagate the local state estimate in time:
\begin{align*}
    \mathbf{P}_i^+ &\leftarrow \mathbf{A} \mathbf{M}_i \mathbf{A}^\top + \mathbf{B} \mathbf{Q} \mathbf{B}^\top,\\
    \quad \bar{\mathbf{x}}_i^+ &\leftarrow \mathbf{A} \hat{\mathbf{x}}_i.
\end{align*}
\end{algorithmic}
\end{algorithm}

\section{SIMULATION RESULTS}

This section evaluates the proposed filter against CKF, KCF, and GKCF using the simulation model in \cite{kamal2013information}. A target moves under stochastic process noise within a region monitored by ten sensor agents, of which at least one is assumed to be oblivious. The state and measurement equations are modeled as LTI systems, as given in \eqref{sys1} and \eqref{meas1}. The state vector $\mathbf{x}\in\mathbb{R}^4$ consists of four elements representing the positions and velocities along two dimensions. The noise covariances are $\mathbf{Q}=\text{diag}(10,10,1,1)$ and $\mathbf{R}_i=100\mathbf{I}_2$, $\forall i$. The process noise input matrix $\mathbf{B}=\mathbf{I}_4$, while the state propagation and measurement matrices are defined as follows:
\begin{align*}
    \mathbf{A} = \begin{bmatrix}
    1 & 0 & 1 & 0\\
    0 & 1 & 0 & 1\\
    0 & 0 & 1 & 0\\
    0 & 0 & 0 & 1
    \end{bmatrix},\ \mathbf{H}_i = \begin{cases} \begin{bmatrix}
    1 & 0 & 0 & 0\\
    0 & 1 & 0 & 0
    \end{bmatrix},\ \text{if}\ i \in \mathcal{D}, \\[1.5em]
    \begin{bmatrix}
    0 & 0 & 0 & 0\\
    0 & 0 & 0 & 0
    \end{bmatrix},\ \text{if}\ i \notin \mathcal{D}.
    \end{cases}
\end{align*}

\begin{figure}
    \centering
    \includegraphics[width=0.4\textwidth]{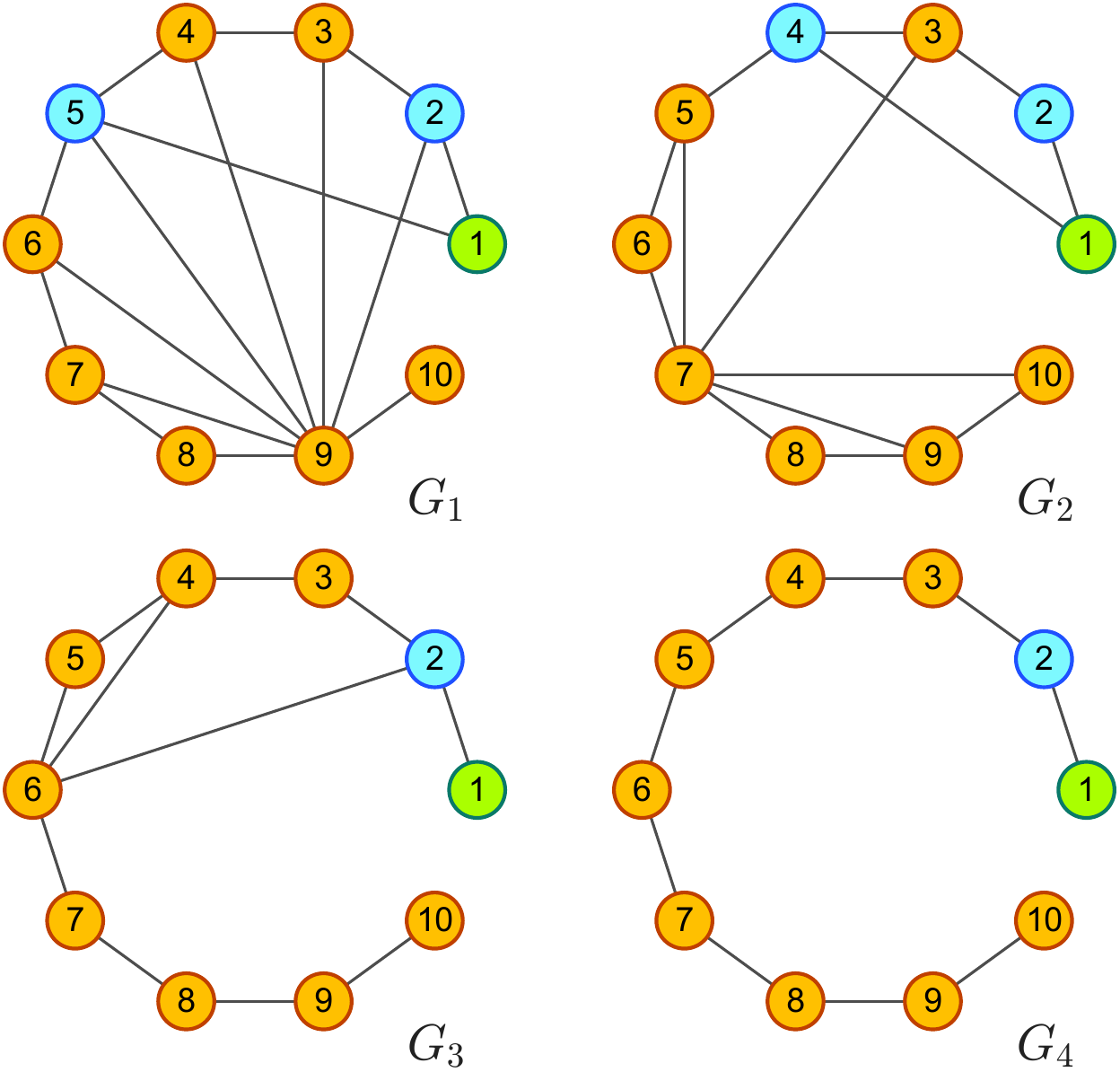}
    \caption{Communication network topologies.}
    \label{fig:network1}
\end{figure}

The consensus step size $\epsilon$ is selected as $0.65/\Delta_{0}$ for the applicable algorithms, where $\Delta_0 = \Delta(G_1)$, with $\mathcal{G}= \{ G_1, G_2, G_3, G_4 \}$. The elements of $\mathcal{G}$ are illustrated in Fig.~\ref{fig:network1}. At $k=150$, the communication graph is switched from $G_1$ to $G_2$, and subsequently to $G_3$ and $G_4$ at $k=300$ and $k=450$, respectively. It is assumed that only a single agent directly observes the target at each time step.

\begin{figure}
    \centering
    \includegraphics[width=0.45\textwidth]{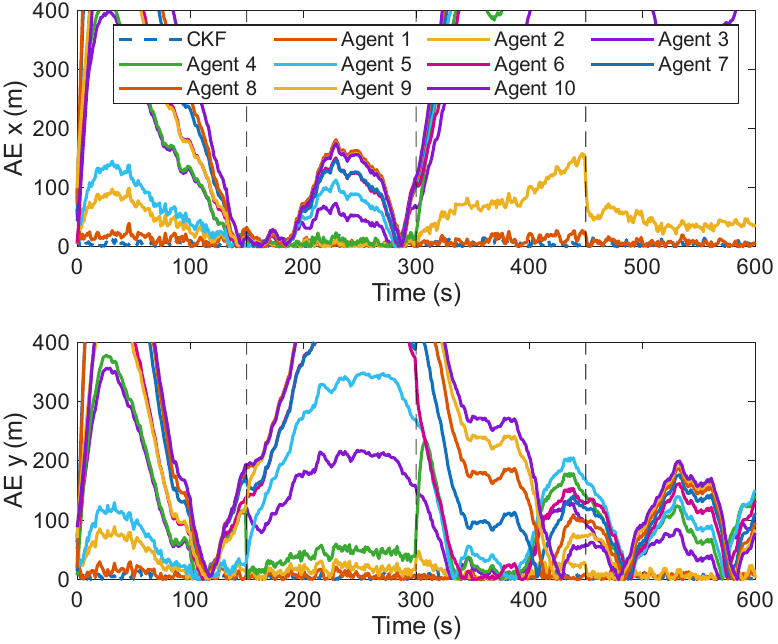}
    \caption{Absolute errors of the KCF agent estimates of target position.}
    \label{fig:ae_kcf}
\end{figure}

\begin{figure}
    \centering
    \includegraphics[width=0.45\textwidth]{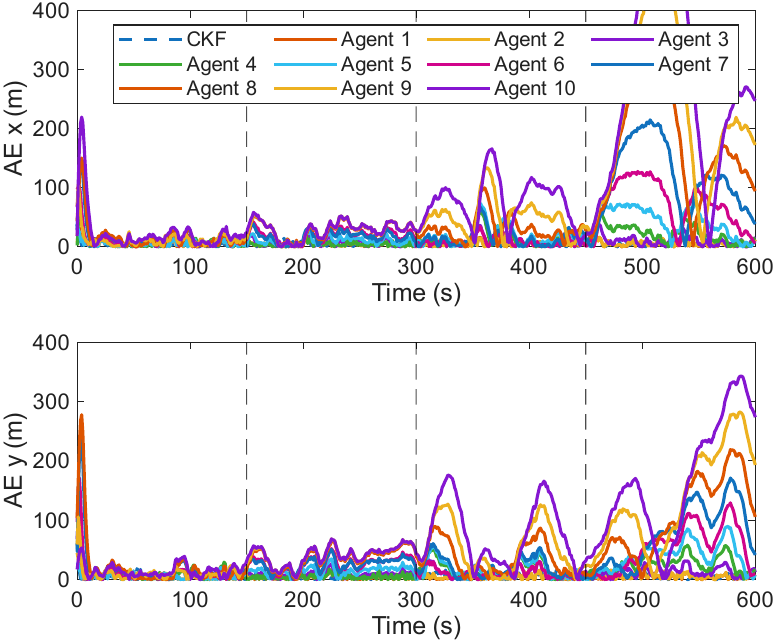}
    \caption{Absolute errors of the GKCF agent estimates of target position.}
    \label{fig:ae_gkcf}
\end{figure}

\begin{figure}
    \centering
    \includegraphics[width=0.45\textwidth]{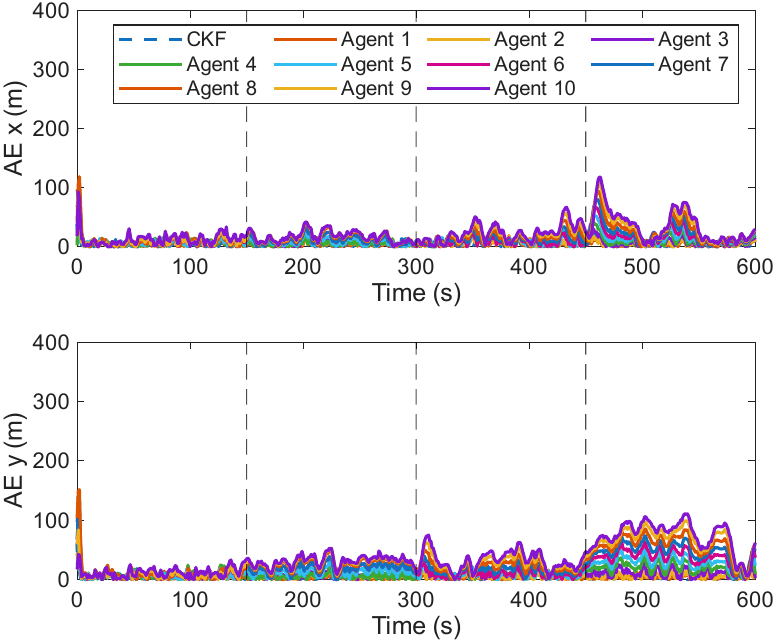}
    \caption{Absolute errors of the DGKCF agent estimates of target position.}
    \label{fig:ae_dgkcf}
\end{figure}

\begin{figure}
    \centering
    \includegraphics[width=0.44\textwidth]{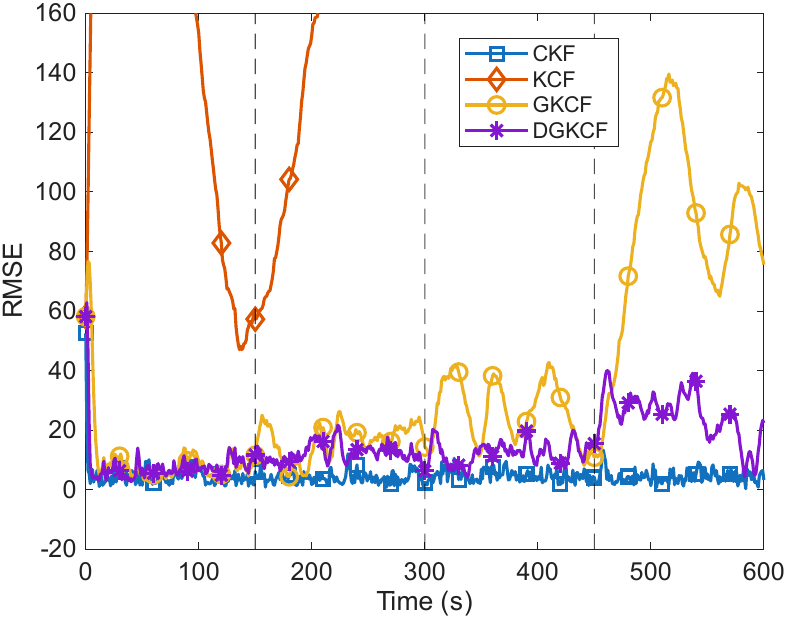}
    \caption{RMSE of the estimated state vectors.}
    \label{fig:rmse}
\end{figure}

The per-agent absolute errors (AE) in position estimates for KCF, GKCF, and DGKCF are shown in Figs.~\ref{fig:ae_kcf}, \ref{fig:ae_gkcf}, and \ref{fig:ae_dgkcf}, respectively. Figure~\ref{fig:rmse} presents the root mean square error (RMSE) of state estimates, averaged over all agents, for KCF, GKCF, and DGKCF, with CKF as a reference. The presence of oblivious agents results in an initially high RMSE for KCF. At $k=150$, the network topology change reduces connectivity, degrading the performance of all distributed filters, with further decline after subsequent switches. Table~\ref{metrics} summarizes the mean RMSE, mean absolute error (MAE), averaged across agents, and the computation time per iteration (computational cost). Additionally, Table~\ref{metrics} presents the time required for the RMSE to first fall below 5. The DGKCF has a computational cost comparable to GKCF while converging faster than GKCF,  and approaching the CKF benchmark, whereas KCF does not converge within the simulation duration.

\renewcommand{\arraystretch}{1.5}
\begin{table}
\caption{Comparison of Performance Metrics}
\label{metrics}
\begin{center}
\begin{tabular}{|l|r|r|c|c|}
\hline
Filter & RMSE & MAE & Comp. Cost (ms) & Conv. Time (s)\\
\hline \hline
CKF & 4.49 & 3.64 & $0.0641$ & 4\\
\hline
KCF & 363.68 & 177.94 & $0.2383$ & ---\\
\hline
GKCF & 35.10 & 18.38 & $0.4286$ & 17\\
\hline
DGKCF & 14.42 & 9.76 & $0.4323$ & 9\\
\hline
\end{tabular}
\end{center}
\end{table}

After each graph switch, KCF and GKCF perform worse than DGKCF because the initially chosen consensus step size is no longer optimal for the updated network topologies. This results in slow convergence of the oblivious agent estimates of these filters, which depend on a properly tuned consensus step size for satisfactory performance.

\section{CONCLUSION}

This paper presented the Dynamic Generalized Kalman Consensus Filter (DGKCF), a distributed estimation algorithm for target tracking in sensor networks with limited sensing ranges and switching communication topologies. The proposed method replaces the fixed consensus step size used in conventional weighted average consensus schemes with locally computed information-based weights derived from the prior error covariance matrices of neighboring agents. Since the consensus weights are computed using only locally available information, the proposed approach requires no global network parameters and can adapt to changes in network connectivity. Simulation results for a target tracking problem demonstrated that DGKCF achieves lower RMSE and MAE, as well as faster convergence, than KCF and GKCF under switching topologies.

Future work will focus on establishing formal convergence guarantees for the proposed algorithm and extending the framework to multi-target tracking applications.


\bibliographystyle{IEEEtran}
\bibliography{IEEEabrv, mybibfile}

\end{document}